\documentclass{article}

\usepackage{cite}
\usepackage{amsmath,amssymb,amsfonts}
\usepackage{algorithm, algorithmic}
\usepackage[dvipdfmx]{graphicx}
\usepackage{textcomp}
\usepackage{xcolor}
\usepackage{multirow}
\usepackage{url}
\usepackage{authblk}

\begin{document}
\title{Heterogeneous computing in a strongly-connected CPU-GPU environment: fast multiple time-evolution equation-based modeling accelerated using data-driven approach}

\author[1]{Tsuyoshi Ichimura}
\author[1]{Kohei Fujita}
\author[2]{Muneo Hori}
\author[1]{Maddegedara Lalith}
\author[3]{Jack Wells}
\author[3]{Alan Gray}
\author[3]{Ian Karlin}
\author[3]{John Linford}
\affil[1]{Earthquake Research Institute, The University of Tokyo, Japan}
\affil[2]{Research Institute for Value-Added-Information Generation, Japan Agency for Marine-Earth Science and Technology, Japan}
\affil[3]{NVIDIA, USA}
\date{}

\maketitle

\begin{abstract}
We propose a CPU-GPU heterogeneous computing method for solving time-evolution partial differential equation problems many times with guaranteed accuracy, in short time-to-solution and low energy-to-solution. On a single-GH200 node, the proposed method improved the computation speed by 86.4 and 8.67 times compared to the conventional method run only on CPU and only on GPU, respectively. Furthermore, the energy-to-solution was reduced by 32.2-fold (from 9944 J to 309 J) and 7.01-fold (from 2163 J to 309 J) when compared to using only the CPU and GPU, respectively. Using the proposed method on the Alps supercomputer, a 51.6-fold and 6.98-fold speedup was attained when compared to using only the CPU and GPU, respectively, and a high weak scaling efficiency of 94.3\% was obtained up to 1,920 compute nodes. These implementations were realized using directive-based parallel programming models while enabling portability, indicating that directives are highly effective in analyses in heterogeneous computing environments.
\end{abstract}

\section{Introduction}
\label{sct1}

Recent advancements in computational capabilities enabled the use of numerical analysis of time-evolution partial differential equation (PDE) problems with guaranteed accuracy not only for forward analysis but also for a wide variety of other applications such as ensemble evaluation, optimization, inverse problems, and surrogate model training. Therefore, there is a growing demand for reducing the time-to-solution and energy-to-solution of numerical analyses.

Advances in the computing environment such as the increase in CPU memory capacity and advances in GPU computing can help meet these needs. The increased CPU memory capacity per computation node enables the analysis of large amounts of data, allowing the implementation of a data-driven approach to improve the efficiency of typical numerical analysis of PDEs for time-evolution problems. In addition, as the development cost of GPU computing has been reduced by directive-based parallel programming models, GPU computing has become popular, and it has become easier to benefit from the computing performance of GPUs that are higher than that of CPUs for general types of computations. Indeed, GPU computing has become a common approach to improving the efficiency of the numerical analysis of time-evolution PDE problems.

As mentioned above, the efficiency of CPU-based analysis is expected to be improved through the use of data-driven approaches because of the large memory available in CPU computing. However, the computational performance of CPUs is lower than that of GPUs; thus, the improvement in efficiency is limited compared to the use of GPUs. In addition, when using GPUs for computation, since GPUs have high computing performance and the data-transfer cost between the CPU and GPU is relatively high,  it is difficult to improve efficiency on GPUs by introducing data-driven approaches while maintaining the problem size due to the limited memory capacity of GPUs. At the same time, with the development of new computer systems, fast connections between CPU and GPU are being developed (e.g., GH200 \cite{GH200}, PCIe (Gen5)\cite{Gen5}). Furthermore, since computation performance increases, memory capacity and memory bandwidth tend to be relatively reduced. With such development of heterogeneous environments including various computer systems and computing mechanisms, heterogeneous computing methods that can take advantage of these systems are considered a promising solution for improving the performance of numerical analysis.

In this paper, using a general problem setting in which a time-evolution problem is solved many times as a demonstrative example, we propose a heterogeneous computing method that combines the data-driven approach, which leverages large CPU memory, with equation-based modeling, which leverages fast GPU computing, in a strongly-connected CPU-GPU environment. We show that the proposed method can be implemented using directive-based parallel programming models to accelerate analysis and improve energy efficiency while enabling portability. To elucidate the improved performance of the proposed method, emphasize its reproducibility to other problems, and facilitate the development and deployment of other methods, the proposed method was constructed using general operations and simple implementations (i.e., we can expect significant performance improvement as shown in this paper when applying the proposed method to other general computing methods). Although a data-driven approach is used in this proposed method, the accuracy of the analysis is guaranteed to be equivalent to that of standard equation-based modeling because the proposed method includes the refinement process (details of the refinement process are explained in Section~\ref{sct2}).

The rest of this paper is organized as follows. In Section~\ref{sct2}, we describe conventional solution methods for time-evolution PDE problems, elaborate on data-driven approaches used to accelerate such computations, and elucidate the construction of a heterogeneous computing method in a strongly-connected CPU-GPU environment. In Section~\ref{sct3}, we demonstrate the effectiveness of the proposed method for solving multiple cases of dynamic elasticity problems. Section~\ref{sct4} presents a summary of this paper and future perspectives.

\section{Method}
\label{sct2}

\subsection{Target problem \& baseline method}

We propose a heterogeneous computing method for accelerating the solution of many cases  of
\begin{equation}
\label{eq:general}
\mathbf{A}^{it} \mathbf{x}^{it} = \mathbf{f}^{it},
\end{equation}
which is a generalized form of a discretized PDE-based time-evolution problem, through the use of equation-based modeling and a data-driven approach in a strongly-connected CPU-GPU environment. Here, $it$ is the time-step, $\mathbf{A}^{it}$ is a non-diagonal sparse matrix at step $it$, $\mathbf{x}^{it}$ is the solution for step $it$, and $\mathbf{f}^{it}$ is a known vector computed using the outer force at step $it$ and response up to step $it$. This problem is sequentially solved in time from step 1 to step $nt$. We suppose the degrees of freedom of $\mathbf{x}^{it}$ is large.

The approach of storing $\mathbf{A}^{it}$ in memory in a compressed form and using it in iterative solvers is often employed for solving Eq.~\eqref{eq:general}. Therefore, herein, as a baseline method, we used Compressed Row Storage (CRS) \cite{CRS-PCG-BJ} as the compression format and the conjugate gradient method \cite{CRS-PCG-BJ} as the iterative solver method, which we hereafter refer to CRS-CG (see Algorithm \ref{BJCG} for the concrete example). Since this is usually computed using only the CPU or GPU, we refer to this baseline method as CRS-CG@CPU and CRS-CG@GPU when run on the CPU and GPU, respectively (see Algorithm \ref{CRS-CG@CPU/GPU} for the concrete example). Although we used a conjugate gradient solver assuming a symmetric positive definite matrix $\mathbf{A}^{it}$, we can choose a suitable iterative solver depending on the characteristics of $\mathbf{A}^{it}$ and apply the proposed method explained in the following subsection.

\subsection{Proposed heterogeneous computing method}

CRS-CG is a typical numerical simulation method based on equation-based modeling. In contrast, the proposed method aims to improve the time-to-solution and energy-to-solution of computations by realizing heterogeneous computing on heterogeneous systems by exploiting the performance of GPUs for improving equation-based modeling and employing a data-driven approach on CPUs. The details of the proposed method are provided below.

First, we attempted to leverage the computing power of GPU. One of the approaches often used in solving Eq.~\eqref{eq:general} via an iterative method is keeping $\mathbf{A}^{it}$ in memory in a compressed format such as CRS, which is efficient when the memory bandwidth is relatively large compared to the computing performance. However, considering that in recent GPUs, the memory bandwidth is becoming smaller relative to the computing performance, there is room to further exploit computing performance \cite{WACCPD2022}. At the same time, computational performance may be improved by addressing the limited GPU memory capacity and the cost of reconstructing the matrix at each time step in nonlinear problems. Thus, the matrix-vector product $\mathbf{A}^{it} \mathbf{x}^{it}$ within the iterative solver used for solving Eq.~\eqref{eq:general} can be computed as
\begin{equation}
\label{eq:EBE}
\sum_e \mathbf{P}^T_e (\mathbf{A}_e (\mathbf{P}_e \mathbf{x}^{it})),
\end{equation}
to prevent the storage of the matrix in memory and the construction of the matrix at each time step. Here, $\mathbf{A}_e$ indicate a $30\times30$ element matrix, and $\mathbf{P}_e$ indicate a $30\times (3n)$ mapping matrix between global node numbers to the element node numbers, which corresponds to random access in memory. This method (called the Element-by-Element (EBE) method \cite{EBE-FEM} in the finite element method) overlaps core kernels; thus, it is faster than the computation of CRS-based matrix-vector products on systems such as GPUs, where the computing performance is high relative to the memory bandwidth. In addition, as the matrices need not be stored in memory, the use of Eq.~\eqref{eq:EBE} enables faster analysis and requires less memory. The reduction in memory footprint enables the concurrent computation of multiple cases, taking advantage of the extra memory capacity (i.e., a single iterative solver is used to analyze $r$ cases simultaneously). In this case, $r$ random data accesses involved in multiplying $\mathbf{P}^T_e$ and $\mathbf{P}_e$ in the evaluation of Eq.~\eqref{eq:EBE} can be accessed sequentially; thus, better computation performance is expected compared to the case where $r$ cases are computed independently. Note that the introduction of EBE makes the computations matrix-free, enabling the use of the proposed method for solving nonlinear problems and simultaneous calculations of the responses of different numerical analysis models (multiple types of $\mathbf{A}$ can be computed at once). At this stage, fast equation-based computation of $r$ cases, which is difficult to perform using CRS-CG@GPU because of limited memory capacity, can be achieved. Note that although the degrees of freedom and model may differ for each case, similar convergence performance must be ensured to achieve the performance described in Section~\ref{sct3}.

Next, we attempted to introduce a data-driven method. As a considerable portion of the GPU memory capacity is used for iteratively solving $r$-cases concurrently, we employed a data-driven method on the CPU. Many data-driven methods, such as the direct estimation of the simulation behavior using a surrogate model learned from the results of previous simulations, have been previously proposed \cite{HPCAsia2022,DMD}. To achieve numerical accuracy of simulation results, we used a data-driven method to estimate a highly accurate initial solution for an iterative solver for solving Eq.~\eqref{eq:general}, by using the simulation results of many previous time steps  (i.e., the accuracy of the results estimated by a data-driven method is not guaranteed; thus, it is refined by an iterative method such that the accuracy is guaranteed). If a highly accurate initial solution can be obtained using the data-driven method, the number of iterations required for solving Eq.~\eqref{eq:general} can be reduced, thereby accelerating computations while ensuring high accuracy. A conventional data-driven method consists of predicting time-evolution using point-wise data for a small number of time steps; however, a more advanced method of predicting solutions by learning large-scale time series data with many steps has been recently developed to take advantage of improved computer performance and memory capacity. The use of this method is expected to enable the prediction of a more accurate initial solution. For example, a method has been proposed in \cite{HPCAsia2022} to efficiently predict the solution of the next step on a massively parallel computer using the solution of the previous $s$ time steps as an input, resulting in a threefold reduction in both the number of solver iterations and the time-to-solution. These methods can be generally expressed as
\begin{equation}
\bar{\mathbf{x}}^{it} \Leftarrow predictor(\mathbf{X}^{it}, \mathbf{F}^{it}, \mathbf{f}^{it}), \label{eq:predictor}
\end{equation}
where $\mathbf{F}^{it}=\left\{ \mathbf{f}^{it-s}, \mathbf{f}^{it-s+1},\mathbf{f}^{it-s+2},...,\mathbf{f}^{it-1} \right\}$ and $\mathbf{X}^{it}=\left\{ \mathbf{x}^{it-s}, \mathbf{x}^{it-s+1},\mathbf{x}^{it-s+2},...,\mathbf{x}^{it-1} \right\}$ are the inputs and outputs of the previous $s$ time steps stored in memory, respectively. If the degrees of freedom of the unknown vectors is $n$, the data size to be stored is proportional to $n \times s$; thus, a considerable amount of memory is required to store $s$ steps needed for the accurate prediction of the initial solution, making it difficult to perform such computations on a GPU. In this case, the input for the predictor ($\mathbf{f}^{it}$) and the prediction result ($\bar{\mathbf{x}}^{it}$), as well as the analysis result $\mathbf{x}^{it}$ used for learning, must be passed between the GPU and CPU, which often becomes a bottleneck in conventional GPU computing. At the same time, the training and prediction costs on the CPU are non-negligible compared to the iterative solver. Thus, the data transfer cost between the CPU and GPU, as well as the training and prediction costs on the CPU become the computational bottleneck of this approach. 

Therefore, we propose a method for solving PDE-based time-evolution problems via simultaneous computing on CPUs and GPUs in a recently developed, strongly connected CPU-GPU environment. Herein, two sets of PDE-based time-evolution problems of similar sizes are simultaneously solved. While the iterative solver for one of the problems is computed on the GPU (hereafter referred to as the solver), the learning/prediction of the solution for the other problem is computed on the CPU (hereafter referred to as the predictor). Once the solver and the predictor computations are complete, the solution and prediction results are quickly synchronized using the fast interconnection between the CPU and GPU. EBE is used in the solver, each using an iterative solution method for solving $r$-cases concurrently, resulting in the computation of 2 sets $\times r$ cases $= 2r$ problem cases of simulations in a single analysis. Here, two processes are run on the same compute node.  The CPU and GPU are used simultaneously in the predictor and solver, and data is passed between the CPU and GPU before and after the computation. To prevent competition between multiple processes using the GPU simultaneously, the GPU is used only by one process at a time. If the execution time of the predictor on the CPU and the execution time of the solver on the GPU are nearly equal, the CPU and GPU computations can be completely overlapped. For example, if the number of solver iterations is reduced by three times as a result of the use of a data-driven method, as in the case of \cite{HPCAsia2022}, the predictor can be hidden using a strongly-connected CPU-GPU environment, and the analysis time can be reduced by three times relative the conventional method. As the computational cost of the predictor depends on the number of time steps $s$ used in $\mathbf{X}^{it}$ and $\mathbf{F}^{it}$ in Eq.~\eqref{eq:predictor}, $s$ is adjusted automatically during the time-history analysis to balance the computation times of the predictor on the CPU and the solver on the GPU.

Hereafter, the proposed heterogeneous computing method is referred to as EBE-MCG@CPU-GPU; it leverages both the high performance of GPUs and the large memory capacity of CPUs in a heterogeneous system with a strongly connected CPU-GPU environment (see Algorithm \ref{EBE-MCG@CPU-GPU} for the concrete example).

Although EBE is effective as described above, for some problems, its implementation may be difficult to achieve high performance such as that described in Section~\ref{sct3}. In such cases, the proposed heterogeneous computing method can be configured using CRS instead of EBE, which is hereafter denoted as CRS-CG@CPU-GPU (see Algorithm \ref{CRS-CG@CPU-GPU} for the concrete example). However, in this case, as shown in the performance measurement in Section~\ref{sct3}, the performance is limited than that of EBE-MCG@CPU-GPU because of the large memory usage for CRS storage and the difficulty of fully utilizing the GPU's computing performance. Note that CRS-CG@CPU-GPU corresponds to the addition of a CPU-based predictor to CRS-CG@GPU and the overlapping of the CPU and GPU execution times.

\section{Numerical Experiment}
\label{sct3}

In this section, we present a concrete form of the proposed method and measure its performance. Although the proposed method can be applied to nonlinear problems (which is another advantage of the matrix-free EBE-MCG@CPU-GPU over the CRS-based method), we use an ensemble simulation of a linear dynamic elastic problem attributed to Eq.~\eqref{eq:general} as an example to demonstrate the differences in performance between the proposed and baseline methods. Specifically, we show the concrete forms of CRS-CG@CPU, CRS-CG@GPU, EBE-MCG@CPU-GPU, and CRS-CG@CPU-GPU, and compare their performances to demonstrate the effectiveness of the proposed method. 

\subsection{Problem setting \& results of application example}
\label{sct3a}

We consider solving many cases of the time-evolution problem of a linear elastic body $V$ 
\begin{equation}
\label{GE:ORG}
\rho \ddot{\mathbf{u}}-(\nabla \cdot \mathbf{c} \cdot \nabla) \cdot \mathbf{u}=\mathbf{f},
\end{equation}
to analyze the response to random wave inputs (random input wave settings are different in each case). Here, $\rho$, $\mathbf{u}$, $\mathbf{c}$, $\mathbf{f}$ indicate the density, the displacement, the elasticity tensor, and outer force, respectively, and $(\dot{~})$, $\nabla$ are the temporal and spatial differential operators. Although there are many uses for such an analysis, we consider a general problem of estimating the properties of $V$ using the obtained random response as described below.

The problem setting shown below is based on 3D ground structure estimation. That is, by solving Eq.~\eqref{GE:ORG} on a 3D ground structure model many times and processing the ensemble responses, we obtain the dominant frequency at each point on the ground surface. By comparing the computed dominant frequency with that obtained from the observed microtremors (constant vibration of the ground) at actual sites, we evaluate the credibility of the 3D ground structure model. The spatiotemporal discretization often used in such analyses is applied to Eq.~\eqref{GE:ORG}; specifically, we obtain the equation below by discretizing space by second-order tetrahedral elements (one of the finite elements), applying the Newmark's $\beta$ method \cite{newmark} (a type of time integration), and considering damping:
\begin{equation}
\left(\frac{1}{dt^2}\mathbf{M} + \frac{1}{dt}\mathbf{C} + \mathbf{K} \right) \mathbf{u}^{it} =
\mathbf{f}^{it} + \mathbf{C}  \mathbf{v}^{it-1} +\mathbf{M} \left( \mathbf{a}^{it-1} + \frac{4}{dt}\mathbf{v}^{it-1} \right).
\label{GE:DIS}
\end{equation}
Here, $dt$ is the time increment width, $\mathbf{M}$, $\mathbf{C}$, $\mathbf{K}$ are the mass, damping, and stiffness matrices, and $\mathbf{u}^{it}, \mathbf{v}^{it}, \mathbf{a}^{it}, \mathbf{f}^{it}$ are the displacement, velocity, acceleration, and outer force vectors at time-step $it$, respectively. Using $\mathbf{u}^{it}$ obtained by solving Eq.~(\ref{GE:DIS}), we obtain
\begin{eqnarray}
\label{timestep}
\mathbf{v}^{it} &\Leftarrow& -\mathbf{v}^{it-1} +  \frac{2}{dt} \left( \mathbf{u}^{it} - \mathbf{u}^{it-1} \right), \\
\mathbf{a}^{it} &\Leftarrow& -\mathbf{a}^{it-1} +  \frac{4}{dt} \mathbf{v}^{it-1} + \frac{4}{dt^2} \left(  \mathbf{u}^{it} - \mathbf{u}^{it-1} \right),
\end{eqnarray}
and proceed to the next time step ($it+1$). Random waves are analyzed by inputting impulse waveforms with random amplitudes and uniform spectra in random directions at 10,000 randomly selected points on the ground surface, and the response is computed for 16384 time-steps with $dt=0.005$. To account for the semi-infinity of the ground, absorbing boundary conditions are applied to the sides of $V$, and the displacement at the bottom is fixed. 

\begin{figure}[tb]
\centering
\includegraphics[width=\hsize]{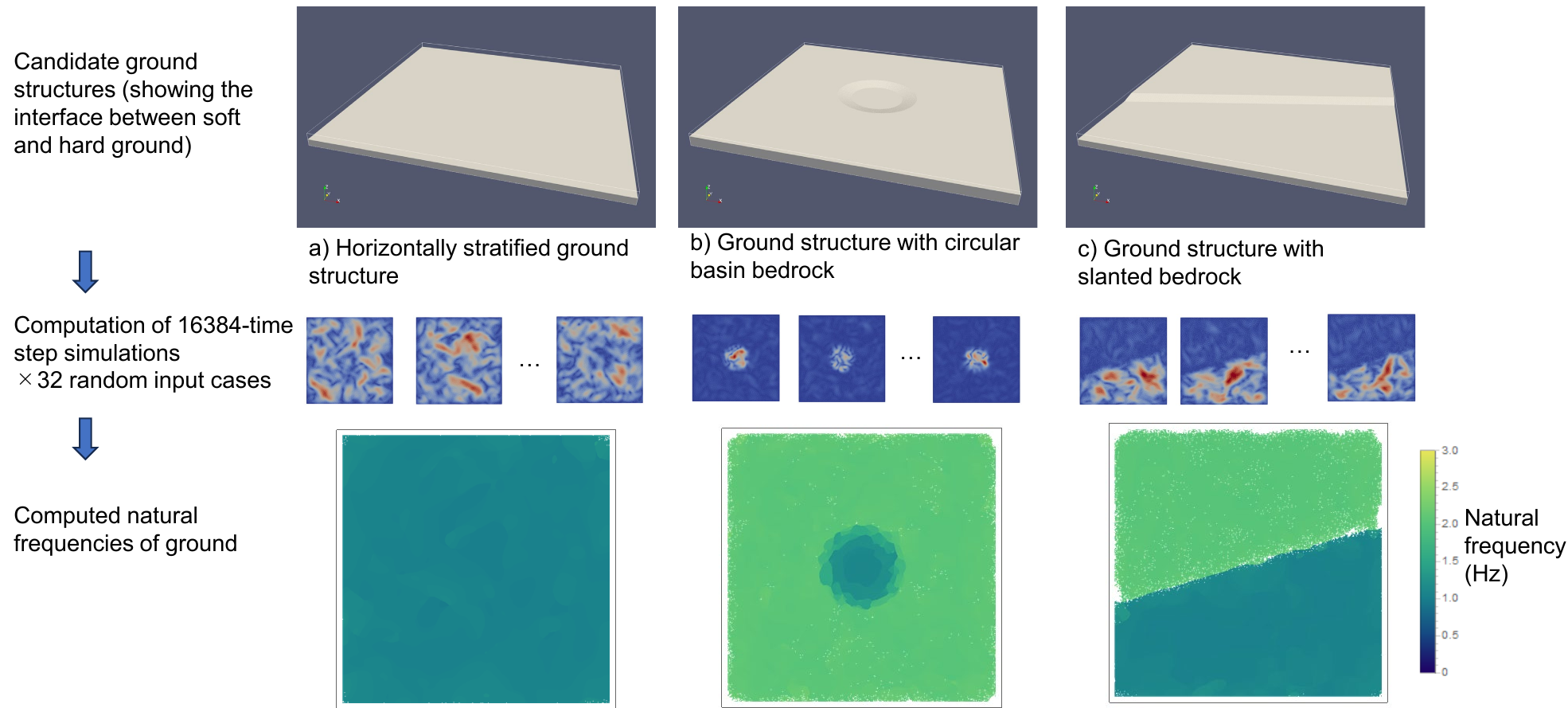}
\caption{Target ground structure and results of frequency domain decomposition. All ground structures have a flat surface but different interface shapes between the sedimentary layer and bedrock. All models have dimensions of 950$\times$950$\times$120 m with a minimum element size of 2.5 m for resolving the frequency components up to 5.0 Hz. The number of second-order tetrahedral nodes and elements in model a are 15,509,903 and 11,365,697, respectively (the number of unknowns in Eq.~\eqref{GE:DIS} is 46,529,709).}
\label{figapplication}
\end{figure}

Free vibration is simulated for 32 random input cases for each of the three types of ground structures shown in Fig.~\ref{figapplication}. The dominant frequency at each point was obtained (the distribution of dominant frequencies obtained for each model is shown in Fig.~\ref{figapplication}) by applying frequency domain decomposition (FDD) \cite{FDD} to the waveforms obtained at each point on the ground surface. All 3D ground structure models exhibit distinct distributions of dominant frequencies. Thus, a candidate 3D ground structure model can be constructed for the site, and its credibility can be evaluated by comparing the distribution of the dominant frequency obtained from the above ensemble simulation with that obtained from microtremor observations. Such analyses are in great demand because the reliability of the 3D ground model is important for designing structures at the site. At the same time, the cost of multiple such analyses is large, not only for the considered example but for other problems as well. Thus, the development of an efficient analysis method, such as the proposed method, is desired.

\newcommand{\MM}{\mathbf{M}}
\newcommand{\MC}{\mathbf{C}}
\newcommand{\MD}{\mathbf{D}}
\newcommand{\MK}{\mathbf{K}}
\newcommand{\MF}{\mathbf{F}}
\newcommand{\MQ}{\mathbf{Q}}
\newcommand{\Mu}{\mathbf{u}}
\newcommand{\Mv}{\mathbf{v}}
\newcommand{\Ma}{\mathbf{a}}
\newcommand{\MA}{\mathbf{A}}
\newcommand{\MB}{\mathbf{B}}
\newcommand{\MG}{\mathbf{G}}
\newcommand{\MN}{\mathbf{N}}
\newcommand{\Mn}{\mathbf{n}}
\newcommand{\Mf}{\mathbf{f}}
\newcommand{\Mr}{\mathbf{r}}
\newcommand{\Mz}{\mathbf{z}}
\newcommand{\Mp}{\mathbf{p}}
\newcommand{\Mq}{\mathbf{q}}
\newcommand{\Mx}{\mathbf{x}}
\newcommand{\My}{\mathbf{y}}
\newcommand{\Me}{\mathbf{e}}
\newcommand{\MP}{\mathbf{P}}
\newcommand{\MX}{\mathbf{X}}
\newcommand{\MY}{\mathbf{Y}}
\newcommand{\MU}{\mathbf{U}}
\newcommand{\Mc}{\mathbf{c}}
\newcommand{\Malpha}{\mathbf{\alpha}}
\newcommand{\Msigma}{\mathbf{\sigma}}
\newcommand{\Mbeta}{\mathbf{\beta}}

\subsection{Concrete form of CRS-CG@CPU, CRS-CG@GPU, EBE-MCG@CPU-GPU, and CRS-CG@CPU-GPU}
\label{sct3b}

To evaluate the performance of the proposed and baseline methods on an ensemble simulation for 32 random input cases on a horizontally stratified model shown in Section~\ref{sct3a}, we first derive the concrete forms of CRS-CG@CPU, CRS-CG@GPU, EBE-MCG@CPU-GPU, and CRS-CG@CPU-GPU as applied to this problem.

The matrix $\left(\frac{1}{dt^2}\mathbf{M} + \frac{1}{dt}\mathbf{C} + \mathbf{K} \right)$ in Eq.~\eqref{GE:DIS} is a positive definite matrix, with three degrees of freedom corresponding to the $x,y,z$ directions per node. When solving such problems, the conjugate gradient method with a 3$\times$3 block Jacobi preconditioner, with the target matrix stored in memory, is often used \cite{CRS-PCG-BJ} (Algorithm~\ref{BJCG}). The main computational cost of such a solver comprises the calculation of the sparse matrix-vector product at each iteration. In this performance measurement, the 3$\times$3 block CRS format \cite{CRS-PCG-BJ}, which is a standard method for storing matrices in memory, is used, and the sparse matrix-vector product is computed by loading the matrix from memory and multiplying it by the right-hand side vector. These computations were performed on the CPU using CRS-CG@CPU and on the GPU using CRS-CG@GPU (Algorithm~\ref{CRS-CG@CPU/GPU}). Herein, the Adams-Bashforth method, which uses the data of four previous time steps, is employed to estimate the initial solution of the solver:
\begin{equation}
\bar{\mathbf{u}}^{it} \Leftarrow \mathbf{u}^{it-1} + 
\frac{dt}{24}\left( -9 \mathbf{v}^{it-4} + 37 \mathbf{v}^{it-3} -59 \mathbf{v}^{it-2} + 55 \mathbf{v}^{it-1} \right).
\nonumber
\end{equation}
As the data used for the Adams-Bashforth method is small, it is stored in the CPU memory in CRS-CG@CPU and in the GPU memory in CRS-CG@GPU.

To further exploit the computing performance of GPU and reduce memory footprint, the sparse matrix-vector product ($\Mq \Leftarrow \MA \Mp$ in Algorithm~\ref{BJCG} line 12), which accounts for the majority of the computational cost in Algorithm~\ref{BJCG}, is changed to a matrix-free operation using the Element-By-Element method \cite{EBE-FEM}:
\begin{equation}
\Mq \Leftarrow \sum_e \MP_e^T \left( \left(\frac{1}{dt^2}\MM_e+\frac{1}{dt}\MC_e + \MK_e \right) \left(\MP_e \Mp \right) \right) \label{eqebefem}
\end{equation}
Here, $\MP_e$ is the mapping matrix between the global nodal number and the local nodal number at element $e$ and $\MM_e, \MC_e, \MK_e$ are the element mass matrix, element damping matrix, and element stiffness matrix, respectively. Hereafter, this method is referred to as EBE-CG@GPU. By applying EBE-CG@GPU to multiple inputs $\Mf$ and combining it with a data-driven method, we construct EBE-MCG@CPU (Algorithm~\ref{EBE-MCG@CPU-GPU}). Owing to reduced memory usage, $r=4$ cases are conducted per set (2 sets $\times$4 cases = 8 problem cases per analysis) in EBE-MCG@CPU-GPU.  In this case, the sparse matrix-vector product in Eq.~\eqref{eqebefem} is transformed to
\begin{equation}
\left\{\Mq_0,\Mq_1,\Mq_2,\Mq_3 \right\} \Leftarrow 
\sum_e \MP_e^T \left( \left(\frac{\MM_e}{dt^2}+\frac{\MC_e}{dt} + \MK_e \right) \left(\MP_e \left\{\Mp_0,\Mp_1,\Mp_2,\Mp_3 \right\} \right) \right). \label{eqmebefem}
\end{equation}
As the random access (e.g., $\MP_e \Mp$ in Eq.~\eqref{eqebefem}) is replaced by block random access (e.g., $\MP_e \left\{\Mp_0,\Mp_1,\Mp_2,\Mp_3 \right\}$), the random access is reduced to $1/r$, and the execution efficiency is expected to be improved.

To examine the performance of the proposed method when the implementation of the above EBE is difficult (equivalent to constructing the method using only the conventional method), we constructed CRS-CG@CPU-GPU shown in Algorithm~\ref{CRS-CG@CPU-GPU}. CRS-CG@CPU-GPU computes the matrix-vector product using 3$\times$3 block CRS and simultaneously computes the predictor for one case on the CPU and the solution of Eq.~\eqref{GE:DIS} for the other case on the GPU. 

\begin{algorithm}[tb]
\caption{\small{ Conjugate gradient method for solving linear set of equations $\mathbf{A} \mathbf{x}=\mathbf{f}$ ($\mathbf{x} \Leftarrow $ CRS-CG($\mathbf{A}$, $\mathbf{f}$, $\bar{\mathbf{x}}$)). $\mathbf{A}$ is stored in CRS in memory, and a 3$\times$3 block Jacobi preconditioner is used. This solver is referred to CRS-CG in the paper. $\bar{\Mx}$, $\MB$, and $\epsilon$ are the initial solution, a 3$\times$3 block Jacobi matrix, and the relative error tolerance. }}
\label{BJCG}
\begin{algorithmic}[1]
\small{
    \STATE $\Mx \Leftarrow \bar{\Mx}$
    \STATE $\Mr \Leftarrow \Mf - \MA \Mx$
    \STATE $\beta \Leftarrow 0$
    \STATE $i \Leftarrow 1$ 
    \WHILE{$\|\Mr\|_2 / \|\Mf\|_2 \ge \epsilon $}
    \STATE $\Mz \Leftarrow \MB^{-1} \Mr$
    \STATE $\rho_a \Leftarrow (\Mz, \Mr)$
    \IF{$i > $ 1}
      \STATE $\beta \Leftarrow \rho_a/\rho_b$
    \ENDIF
    \STATE $\Mp \Leftarrow \Mz + \beta \Mp$
    \STATE $\Mq \Leftarrow \MA \Mp$
    \STATE $\alpha \Leftarrow \rho_a/(\Mp, \Mq)$
    \STATE $\rho_b \Leftarrow \rho_a$
    \STATE $\Mr \Leftarrow \Mr - \alpha \Mq$
    \STATE $\Mx \Leftarrow \Mx + \alpha \Mp$
    \STATE $i \Leftarrow i +1$
    \ENDWHILE
}
\end{algorithmic}
\end{algorithm}

\begin{algorithm}[tb]
\caption{\small{ CRS-CG@CPU and CRS-CG@GPU. ``foo'' in the listing below becomes CPU and GPU for CRS-CG@CPU and CRS-CG@CPU, respectively. The response for nt time steps is obtained sequentially by estimating the initial solution $\bar{\mathbf{x}}^{it}$ using the four previous time steps ($\mathbf{x}^{*}$) by the Adams-Bashforth method, and $\mathbf{A } \mathbf{x}^{it}=\mathbf{f}^{it}$ is solved by CRS-CG using this initial solution.}}
\label{CRS-CG@CPU/GPU}
\begin{algorithmic}[1]
\small{
\FOR{it $\le$ nt}
\STATE $\bar{\mathbf{x}}^{it} \Leftarrow $ Adams-Bashforth($\mathbf{x}^{*}$)@foo ! Predictor \\
\STATE $\mathbf{x}^{it} \Leftarrow $ CRS-CG($\mathbf{A}$, $\mathbf{f}^{it}$, $\bar{\mathbf{x}}^{it}$)@foo ! Solver \\
\ENDFOR
}
\end{algorithmic}
\end{algorithm}

\begin{algorithm}[ptb]
\caption{\small{EBE-MCG@CPU-GPU. Simultaneous CPU and GPU execution enables fast computation. Here, two processes are used and 4 cases are analyzed per process (subscripts $i=0$--$7$ denote case numbers). The solution $\bar{\mathbf{x}}^{it}_i$ is predicted on the CPU using the data of past $s$ time-steps $\mathbf{x}^{*}_i$, which is used as the initial solution for the EBE-CG on the GPU. EBE-CG is a solver that replaces the CRS-based matrix-vector product $\MA \Mp$ in CRS-CG (Algorithm~\ref{CRS-CG@CPU/GPU} line 12) by the Element-by-Element method. The four sets of EBE-CG (lines 9--12 and 18--21) are computed simultaneously, which reduces the random access in the EBE-based sparse-matrix vector product by fourfold compared to the case where they are computed separately. }}
\label{EBE-MCG@CPU-GPU}
\begin{algorithmic}[1]
\small{
\FOR{it $\le$ nt}
\STATE process synchronization
\STATE (CPU side, process \#0) \\
\STATE ~~ $\bar{\mathbf{x}}^{it}_0 \Leftarrow$ Data-driven($\mathbf{x}^{*}_0$)@CPU ! Predictor \\
\STATE ~~ $\bar{\mathbf{x}}^{it}_1 \Leftarrow$ Data-driven($\mathbf{x}^{*}_1$)@CPU ! Predictor \\
\STATE ~~ $\bar{\mathbf{x}}^{it}_2 \Leftarrow$ Data-driven($\mathbf{x}^{*}_2$)@CPU ! Predictor \\
\STATE ~~ $\bar{\mathbf{x}}^{it}_3 \Leftarrow$ Data-driven($\mathbf{x}^{*}_3$)@CPU ! Predictor \\
\STATE (GPU side, process \#1) \\
\STATE ~~ $\mathbf{x}^{it}_4 \Leftarrow$ EBE-CG($\mathbf{A}$, $\mathbf{f}^{it}_4$, $\bar{\mathbf{x}}^{it}_4$)@GPU ! Solver \\
\STATE ~~ $\mathbf{x}^{it}_5 \Leftarrow$ EBE-CG($\mathbf{A}$, $\mathbf{f}^{it}_4$, $\bar{\mathbf{x}}^{it}_5$)@GPU ! Solver \\
\STATE ~~ $\mathbf{x}^{it}_6 \Leftarrow$ EBE-CG($\mathbf{A}$, $\mathbf{f}^{it}_4$, $\bar{\mathbf{x}}^{it}_6$)@GPU ! Solver \\
\STATE ~~ $\mathbf{x}^{it}_7 \Leftarrow$ EBE-CG($\mathbf{A}$, $\mathbf{f}^{it}_4$, $\bar{\mathbf{x}}^{it}_7$)@GPU ! Solver \\%
\STATE process synchronization
\STATE (process \#0) transfer $\bar{\mathbf{x}}^{it}_0$, ..., $\bar{\mathbf{x}}^{it}_3$ from CPU to GPU 
\STATE (process \#1) transfer ${\mathbf{x}}^{it}_4$, ..., ${\mathbf{x}}^{it}_7$ from GPU to CPU 
\STATE process synchronization
\STATE (GPU side, process \#0) \\
\STATE ~~ $\mathbf{x}^{it}_0 \Leftarrow$ EBE-CG($\mathbf{A}$, $\mathbf{f}^{it}_0$, $\bar{\mathbf{x}}^{it}_0$)@GPU ! Solver \\
\STATE ~~ $\mathbf{x}^{it}_1 \Leftarrow$ EBE-CG($\mathbf{A}$, $\mathbf{f}^{it}_1$, $\bar{\mathbf{x}}^{it}_1$)@GPU ! Solver \\
\STATE ~~ $\mathbf{x}^{it}_2 \Leftarrow$ EBE-CG($\mathbf{A}$, $\mathbf{f}^{it}_2$, $\bar{\mathbf{x}}^{it}_2$)@GPU ! Solver \\
\STATE ~~ $\mathbf{x}^{it}_3 \Leftarrow$ EBE-CG($\mathbf{A}$, $\mathbf{f}^{it}_3$, $\bar{\mathbf{x}}^{it}_3$)@GPU ! Solver \\
\STATE (CPU side, process \#1) \\
\STATE ~~ $\bar{\mathbf{x}}^{it+1}_4 \Leftarrow$ Data-driven($\mathbf{x}^{*}_4$)@CPU ! Predictor \\
\STATE ~~ $\bar{\mathbf{x}}^{it+1}_5 \Leftarrow$ Data-driven($\mathbf{x}^{*}_5$)@CPU ! Predictor \\
\STATE ~~ $\bar{\mathbf{x}}^{it+1}_6 \Leftarrow$ Data-driven($\mathbf{x}^{*}_6$)@CPU ! Predictor \\
\STATE ~~ $\bar{\mathbf{x}}^{it+1}_7 \Leftarrow$ Data-driven($\mathbf{x}^{*}_7$)@CPU ! Predictor \\
\STATE process synchronization
\STATE (process \#0) transfer ${\mathbf{x}}^{it}_0$, ..., ${\mathbf{x}}^{it}_3$ from GPU to CPU 
\STATE (process \#1) transfer $\bar{\mathbf{x}}^{it+1}_4$, ..., $\bar{\mathbf{x}}^{it+1}_7$ from CPU to GPU %
\STATE process synchronization
\ENDFOR
}
\end{algorithmic}
\end{algorithm}

\begin{algorithm}[tb]
\caption{\small{CRS-CG@CPU-GPU. Simultaneous CPU and GPU execution enables fast computation. Here, one analysis case is computed for each of the two processes (subscripts $i=0$--$1$ denote case numbers). Solution $\bar{\mathbf{x}}^{it}_i$ is predicted on the CPU using the data for past $s$ time-steps $\mathbf{x}^{*}_i$, which is used as the initial solution for CRS-CG@GPU.}}
\label{CRS-CG@CPU-GPU}
\begin{algorithmic}[1]
\small{
\FOR{it $\le$ nt}
\STATE process synchronization
\STATE (CPU side, process \#0) \\
\STATE ~~ $\bar{\mathbf{x}}^{it}_0 \Leftarrow$ Data-driven($\mathbf{x}^{*}_0$)@CPU ! Predictor \\
\STATE (GPU side, process \#1) \\
\STATE ~~ $\mathbf{x}^{it}_1 \Leftarrow$ CRS-CG($\mathbf{A}$, $\mathbf{f}^{it}_1$, $\bar{\mathbf{x}}^{it}_1$)@GPU ! Solver \\
\STATE process synchronization
\STATE (process \#0) transfer $\bar{\mathbf{x}}^{it}_0$ from CPU to GPU
\STATE (process \#1) transfer ${\mathbf{x}}^{it}_1$ from GPU to CPU 
\STATE process synchronization
\STATE (GPU side, process \#0) \\
\STATE ~~ $\mathbf{x}^{it}_0 \Leftarrow$ CRS-CG($\mathbf{A}$, $\mathbf{f}^{it}_0$, $\bar{\mathbf{x}}^{it}_0$)@GPU ! Solver \\
\STATE (CPU side, process \#1) \\
\STATE ~~ $\bar{\mathbf{x}}^{it+1}_1 \Leftarrow$ Data-driven($\mathbf{x}^{*}_1$)@CPU ! Predictor \\
\STATE process synchronization
\STATE (process \#0) transfer ${\mathbf{x}}^{it}_0$ from GPU to CPU 
\STATE (process \#1) transfer $\bar{\mathbf{x}}^{it+1}_1$ from CPU to GPU
\STATE process synchronization
\ENDFOR
}
\end{algorithmic}
\end{algorithm}

Next we explain the details of the data-driven method used in Algorithm~\ref{EBE-MCG@CPU-GPU} and Algorithm~\ref{CRS-CG@CPU-GPU} (this corresponds to the concrete implementation of the data-driven method in Eq.~\eqref{eq:predictor}). We use a method reported in \cite{HPCAsia2022}, which is based on the orthogonal decomposition of past data, because of its high predictor performance and affinity to massively parallel computing. While the Adams-Bashforth method estimates low-order modes in the initial solution with relatively high accuracy, its estimation performance of higher-order modes is relatively poor, resulting in a limited reduction in the number of solver iterations. To address this issue, the proposed method first estimates the solution using the Adams-Bashforth method, then divides the target region into small regions to obtain the residuals of the estimated solution, and predicts the solution ($\mathbf{x}^{it}$) using the data for the previous $s$ steps in each region. As orthogonal decomposition is the main kernel, the computing performance is readily obtained, and the solution of the next step can be predicted without communication between the regions, making the method suitable for both massively parallel computing and multi-core CPU computing within a node. In addition, because it is a general method not limited to a particular PDE, it can be applied to multiple types of time-evolution problems such as wave propagation problems \cite{HPCAsia2022} and viscoelastic analysis \cite{SCALA2022}, and improvement in initial solution accuracy and reduction in solver iterations was obtained on A64FX CPU-based Fugaku \cite{A64FX,Fugaku}. In this paper, we follow the example of these studies and perform the actual prediction as follows. Given a set of inputs $\mathbf{X} = \{ \mathbf{x}_1, \mathbf{x}_2, ..., \mathbf{x}_s \}$ and corresponding outputs $\mathbf{Y} = \{ \mathbf{y}_1, \mathbf{y}_2, ..., \mathbf{y}_s\}$, using the modified Gram Schmidt method, we compute a $s\times s$ upper triangle matrix $\mathbf{U}$ such that matrix $\mathbf{P} = \mathbf{XU}$ becomes an orthonormal basis. Using $\mathbf{P}$, a given input $\mathbf{x}$ can be decomposed as $\mathbf{x}= \mathbf{Pc}+\mathbf{r}$, where $\mathbf{c}$ can be computed as $\mathbf{c} = \mathbf{P}^T \mathbf{x}$. Using this $\mathbf{c}$, the response for $\mathbf{x}$ can be estimated as $\mathbf{y} = \mathbf{YUc} = \mathbf{YUP}^T \mathbf{x} = (\mathbf{YUU}^T\mathbf{X}^T )\mathbf{x}$. Here, we use the difference in the displacement estimation from the Adams-Bashforth predictor and the true displacement ($\mathbf{x} = \mathbf{u}_\mathrm{true} - \mathbf{u}_\mathrm{adams}$) for the current step as $\mathbf{x}$ and the previous step as $\mathbf{y}$. In the proposed method, the simulation is performed while dynamically adjusting the number of steps $s$ used for the predictor, so that the solver computation time on the GPU is equivalent to the predictor computation time on the CPU.

\begin{figure}[tb]
\begin{center}
\includegraphics[width=0.9\hsize]{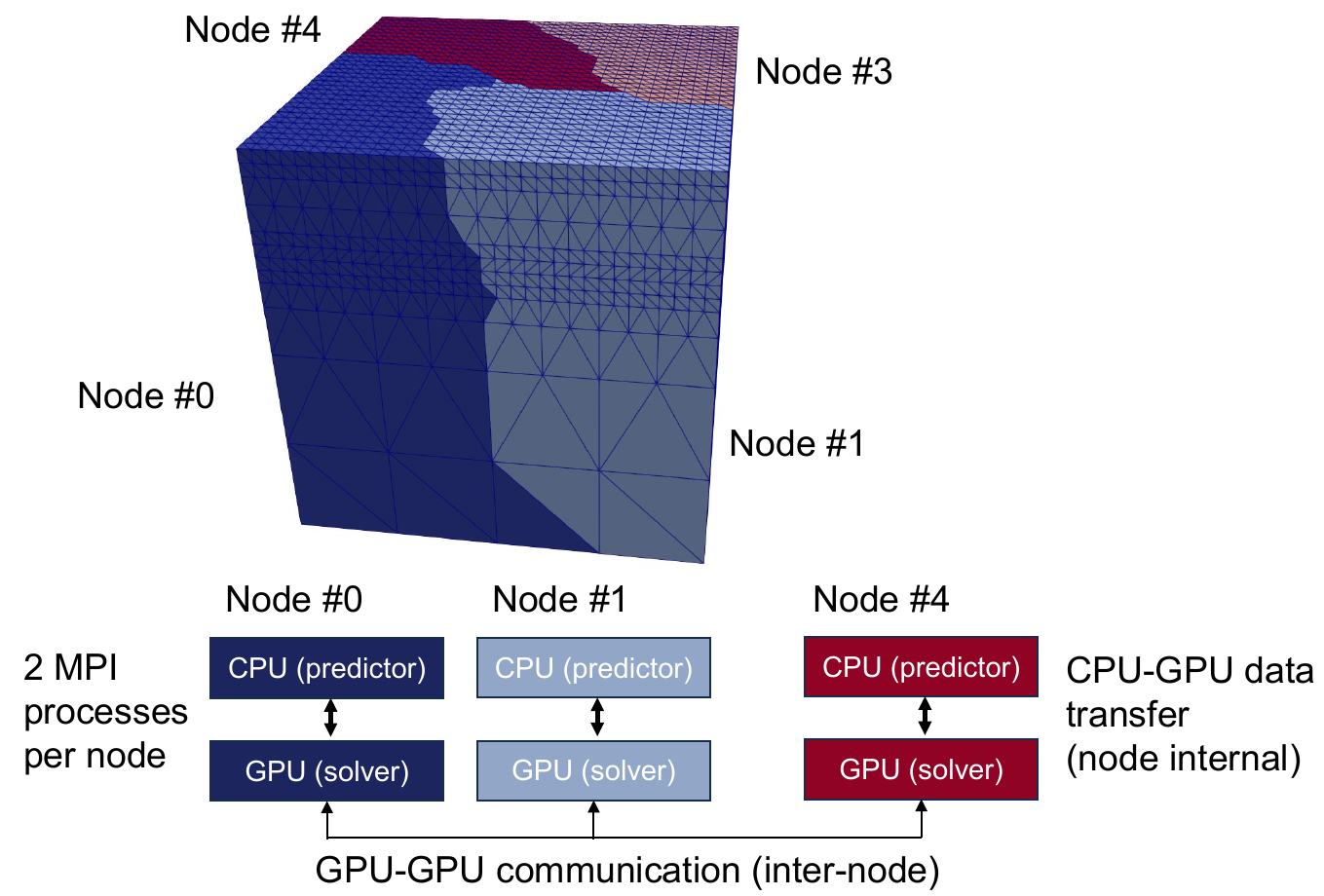}
\end{center}
\caption{Proposed heterogeneous computational algorithm implemented on multiple compute nodes. The finite element model of the target domain is partitioned into the number of compute nodes, and each compute node executes Algorithm~\ref{EBE-MCG@CPU-GPU} using two MPI processes. As the predictor does not require information exchange between partitions, inter-node communication is used only in the solver@GPU, so that the nodal values between partitions are consistent.}
\label{figmultinodediagram}
\end{figure}

As the proposed method consists only of a solver, a data-driven predictor, and data transfer between CPU and GPU, it may be applied to large-scale problems using many compute nodes by selecting scalable methods for the solver and the data-driven predictor.  Herein, the target region is partitioned into the number of compute nodes (or number of GPUs if multiple GPUs are available in each compute node) using a graph partitioning method (e.g., \cite{metis}), and Algorithm~\ref{EBE-MCG@CPU-GPU} is executed on the assigned partition on each compute node (see Fig.~\ref{figmultinodediagram}). The simulation is performed such that the computation becomes consistent with a single CPU-GPU case by conducting point-to-point synchronization between GPUs in EBE-MCG (lines 9--12, 18--21 of Algorithm~\ref{EBE-MCG@CPU-GPU}). As the data-driven predictor does not require the exchange of information between partitions, the parallel performance is not degraded by inter-node communication.

For assessing the general performance of the proposed method on a range of systems within reasonable implementation costs, the programs are implemented using OpenMP \cite{openMP} for multi-core CPU computation, OpenACC \cite{openACC} for GPU computation, and MPI \cite{MPI} for inter-process communication. GPUDirect \cite{GPUDirect} is employed for inter-GPU communication without involving the CPU. Details of the implementation is shown in the AD/AE appendix, where the source codes for the sparse matrix-vector products, which comprises most of the computational cost in the application, are provided.

\subsection{Performance measurement on a single-GH200 node}

As an example of a strongly connected CPU-GPU system, we measure the performance of the proposed method on systems equipped with NVIDIA Grace Hopper Superchip (GH200) \cite{GH200}. As there are multiple versions of GH200 with different specifications, such as memory capacity, we evaluate the robustness of the proposed method by measuring its performance on two versions of GH200. In this section, we report the single-node performance of the version with one CPU and one GPU on a single module (hereafter referred to as single-GH200 node). In Section~\ref{sct3d}, we report the single-node and multiple-node performance of the proposed method on CSCS Alps \cite{ALPS}, which has a different GH200 specification from the single-GH200 node.

Table~\ref{tableenv} lists the specifications of the single-GH200 node. The CPU memory capacity of the single-GH200 node is 480/96 = 5 times larger than its GPU memory capacity. Also, the data transfer bandwidth between the CPU and GPU is 900 GB/s, which is about 1/4 of the GPU memory bandwidth. The power cap of the system is 1,000 W, allowing the CPU cores and the GPU to operate simultaneously at high frequencies. Therefore, the proposed method is expected to run faster than the conventional method because the system has a large CPU memory and high CPU-GPU data transfer bandwidth, and can run both CPU and GPU simultaneously.

\begin{table}[tb]
\caption{Measurement environment. CPU-GPU connection is NVLink-C2C for both systems (900 GB/s bidirectional within each module).}
\label{tableenv}
\centering
\includegraphics[width=\hsize]{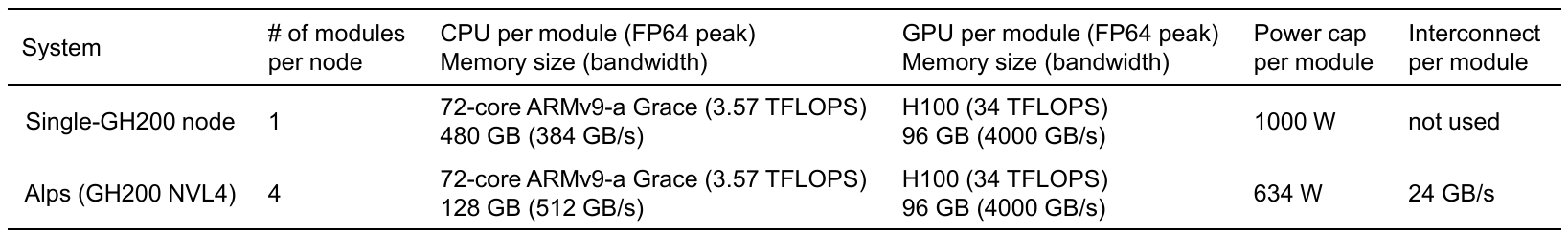}
\end{table}

First, we measure the computational performance of the CRS-based sparse matrix-vector product used in the conventional method and the many-vector version of the EBE-based sparse matrix-vector product used in the proposed method (Table~\ref{tablekernelperformance}). Specifically, we compare the performance of the 3$\times$3 block CRS sparse matrix-vector product kernel, the EBE kernel with one right-hand side (Eq.~\eqref{eqebefem}), and the EBE kernel with multiple right-hand sides (Eq.~\eqref{eqmebefem}). The CRS-based sparse matrix-vector product on the CPU and GPU attain 51.0--54.6\% of the memory bandwidth, indicating that the performance is proportional to the hardware memory bandwidth. Note that this performance is comparable to the block CRS-based matrix-vector product implemented in cuSPARSE \cite{cusparse}; thus, we can consider that the tuning level of the program is at a similar level on both CPU and GPU \cite{WACCPD2022}. The change from CRS to EBE led to a 12.9-fold reduction in memory transfer and a 3.68-fold speedup. Furthermore, the EBE with multiple right-hand sides reduces random accesses, resulting in an additional 1.91-fold speedup. The performance when using CUDA was almost the same as that of OpenACC for this kernel, indicating that the proposed method may be implemented as a portable application while maintaining its performance using directive-based parallel programming models.

\begin{table}[tb]
\caption{Performance of sparse matrix-vector kernel on a single-GH200 node. EBE4 is EBE with $r=4$ right-hand sides.}
\label{tablekernelperformance}
\centering
\begin{tabular}{crrrrr}
\hline
\multirow{2}{*}{Kernel type} & Time & TFLOPS & Mem. bandwidth \\
 & per case & (\% to peak) & TB/s (\% to peak) \\
\hline
CRS-OpenMP@CPU & 163 ms & 0.0485 (1.36\%) & 0.210 (54.6\%) \\
CRS-OpenACC@GPU & 16.8 ms & 0.472 (1.39\%) & 2.04 (51.0\%) \\
EBE-OpenACC@GPU & 4.56 ms & 9.51 (28.0\%) & 0.582 (14.6\%) \\
EBE4-OpenACC@GPU & 2.39 ms & 18.1 (53.3\%) & 0.511 (12.8\%) \\
EBE4-CUDA@GPU & 2.54 ms & 17.1 (50.2\%) & 0.480 (12.0\%) \\
\hline
\end{tabular}
\end{table}

\begin{figure}[tb]
\begin{center}
\includegraphics[width=0.8\hsize]{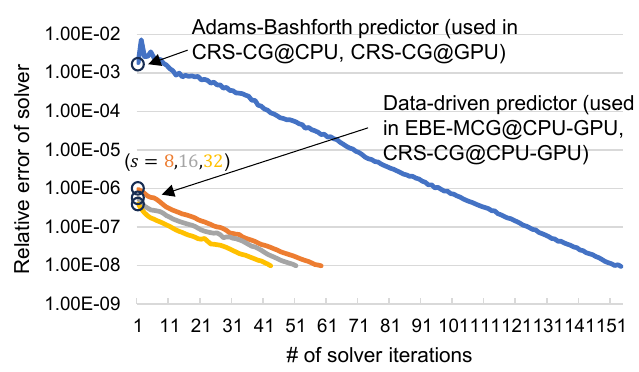}
\end{center}
\caption{Convergence history of the solver for each initial solution estimation method for one time step. Compared to the Adams-Bashforth method used in conventional methods, the number of iterations required to fulfill the error threshold of $\epsilon=10^{-8}$ is reduced by using the data-driven predictor.}
\label{figiterations}
\end{figure}

Next, we confirm the reduction of solver iterations owing to the use of the data-driven method employed in the proposed method. Fig.~\ref{figiterations} shows the convergence history of the error of the solver $\|\Mr\|_2 / \|\Mf\|_2$ with respect to the number of solver iterations for one time step. The use of the data-driven method reduces the error of the initial solution from $1.86\times 10^{-3}$ (Adams-Bashforth method) to $9.46 \times 10^{-7}$, and the number of iterations is correspondingly reduced from 154 to 59 (in case of $s=8$). An increase in the number of steps $s$ used for the predictor improves the accuracy of the initial solution and reduces the number of iterations (iterations are reduced from 59, 51, to 43 by increasing $s$ from 8, 16, to 32). The number of steps $s$ is determined by considering the execution time of the predictor on the CPU and the amount of data that can be stored in memory. In this measurement, the data of the previous 32 steps are stored in CPU memory, and the number of steps $s$ used for the predictor is dynamically selected from the range of $8 \leq s \leq 32$ during the simulation in such a way that the execution time of the predictor@CPU is equivalent to the execution time of the solver@GPU. 

Table~\ref{performance} summarizes the performance measurement results of each method. For the conventional CRS-CG@CPU/GPU method, the average execution time per time step is shown for one case of random input. For the proposed EBE-MCG@GPU-GPU method, responses for eight cases of random inputs are computed, and the average execution time per case and time step is shown. First, we compare the performance of the conventional method running on the CPU only or GPU only (CRS-CG@CPU and CRS-CG@GPU, shown in Algorithm~\ref{CRS-CG@CPU/GPU}). Similar to the performance improvement of the CRS kernel, the use of GPU results in a 9.96-fold speedup, which is close to the memory bandwidth ratio from the CPU. Although the average power consumption increased by 2.17 times (from 327 W for CRS-CG@CPU to 709 W for CRS-CG@GPU), the energy-to-solution was reduced by 4.60 times (from 9944 J to 2163 J) owing to the 9.96-fold reduction in elapsed time (power is measured using ``nvidia-smi -q -d POWER'' \cite{nvidia-smi} and averaged over time). Although the use of the GPU resulted in speedup and reduction in energy-to-solution compared to the use of only CPUs, in this case, CPU cores and CPU memory are almost completely unused; thus, the proposed method is expected to improve the computation speed and energy efficiency.

\begin{figure}[tb]
\centering
\includegraphics[width=0.8\hsize]{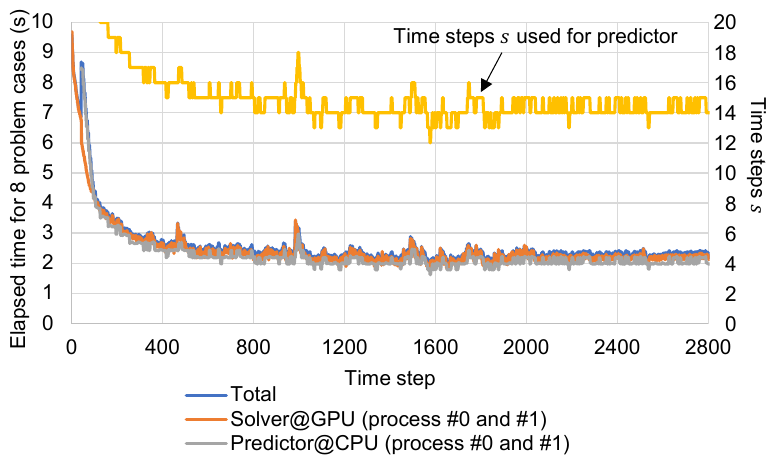}
\caption{Breakdown of elapsed time and selection of $s$ during the simulation in EBE-MCG@CPU-GPU on a single-GH200 node. Although the convergence of the problem changes during the time-history simulation, a suitable $s$ is selected such that the elapsed time of the solver and predictor becomes balanced.}
\label{figstepselection}
\end{figure}

The performance of the baseline method and EBE-MCG@CPU-GPU are compared. 36 CPU cores are used per process to compute the predictor in the proposed method. Fig.~\ref{figstepselection} shows the breakdown of elapsed time and selected $s$ during the simulation. We can see that the number of time steps $s$ used for the predictor is adjusted throughout the simulation such that the elapsed time of the predictor and the solver are almost equal. With the 4.08-fold speedup per solver iteration/case compared to the CRS when using EBE with multiple right-hand sides (Eq.~\eqref{eqmebefem}), the number of iterations per time step reduced from 152 to 68.8 by the introduction of the predictor, and that the execution time of the predictor@CPU is completely hidden by the execution time of EBE-MCG@GPU, we can see that the proposed method shows a 86.4-fold speedup from CRS-CG@CPU and a 8.67-fold speedup from CRS-CG@GPU. While most of the GPU memory (44.9 GB) was used in CRS-CG@GPU and thus additional cases could not be run simultaneously, the introduction of EBE eliminated the need to store the entire matrix in CRS on GPU memory and allowed storing the $2\times 4$ problem cases of simulations to be run simultaneously. The proposed EBE-MCG@CPU-GPU method uses 340 GB from 480 GB of CPU memory, indicating that computations were accelerated by the use of not only the CPU and GPU arithmetic units but also the CPU and GPU memory capacity. As both CPU and GPU are used in EBE-MCG@CPU-GPU, the time-averaged total power consumption of the module including memory, CPU, and GPU is higher (877 W) than those of CRS-CG@CPU (327 W) and CRS-CG@GPU (709 W). However, a substantial reduction in elapsed time resulted in a decrease in energy-to-solution by 32.2 times compared to CRS-CG@CPU (from 9944 J to 309 J) and by 7.01 times compared to CRS-CG@GPU (from 2163 J to 309 J). These results show that effective utilization of each component of a heterogeneous computer system not only shortens execution time but also reduces energy-to-solution.

Lastly, we measured the performance of CRS-CG@CPU-GPU, which corresponds to the proposed method when EBE is not available.
The obtained execution time per solver iteration is equivalent to that of CRS-CG@GPU, but the number of solver iterations is reduced from 152 to 66.6 per time step owing to the use of the data-driven predictor; leading to a 2.61-fold speedup from CRS-CG@GPU. Again, the execution time of the predictor@CPU is overlapped by the computation time of the solver@GPU, indicating that the heterogeneous CPU and GPU computation works effectively. Although in this case, the efficiency is lower than that of EBE-MCG@CPU-GPU, where a highly efficient EBE kernel is used, the energy-to-solution is still significantly lower than that in the baseline method.

\begin{table}[tb]
\caption{Performance of application on a single-GH200 node. Time is shown for the average elapsed time per time step between 250--500th time step simulation per problem case. Module power indicates total power of the module including memory + CPU + GPU. Module and GPU power are time-averaged values.}
\label{performance}
\begin{center}
\includegraphics[width=\hsize]{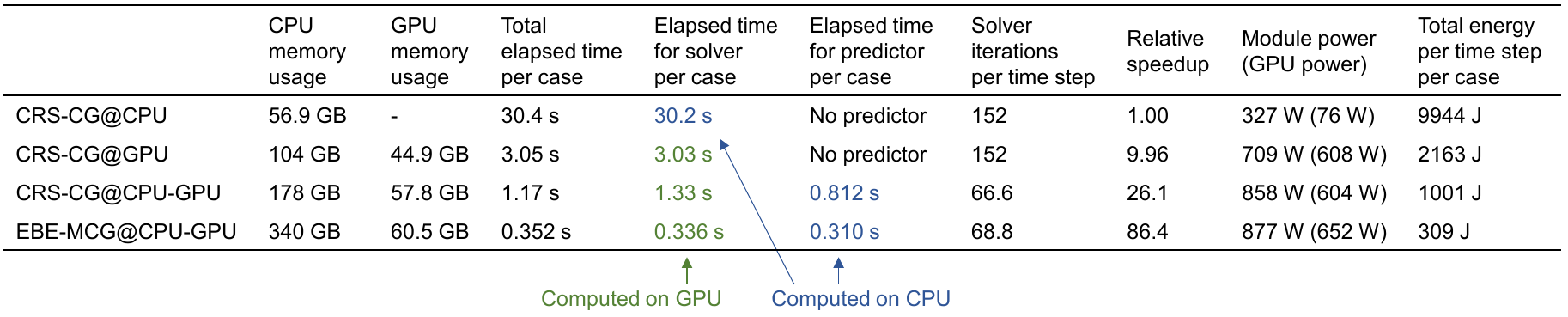}
\end{center}
\end{table}

\subsection{Performance measurement on Alps}
\label{sct3d}

We measure the performance of the proposed method on the GH200 NVL4-based CSCS Alps \cite{ALPS}, which has different memory capacity and power constraints than the single-GH200 node used for the measurements in the previous section. Through measurements on Alps, we show that the proposed method can perform well on systems with different characteristics, requiring only the adjustment of simulation parameters, and that the method is scalable to massively parallel systems.

While the single-GH200 node comprises a single module with one CPU and one GPU, the Alps computation node comprises a NUMA system with four modules, each with one CPU and one GPU (Table~\ref{tableenv}). Its CPU memory bandwidth is higher, but the CPU memory capacity is lower (128 GB per module as opposed to 480 GB in the single-GH200 node). Whereas both the CPU and GPU on the single-GH200 node could be simultaneously run at high loads, the Alps has a power cap of 634 W per module, leading to lower GPU clocks at high CPU loads.

\begin{table}[tb]
\caption{Performance of application on one Alps node. Time is shown for the average elapsed time per time step between 250--500th time step simulation per problem case. Module power indicates the total power of the module including memory + CPU + GPU. Module and GPU power are time-averaged values.}
\label{performanceALPS1}
\begin{center}
\includegraphics[width=\hsize]{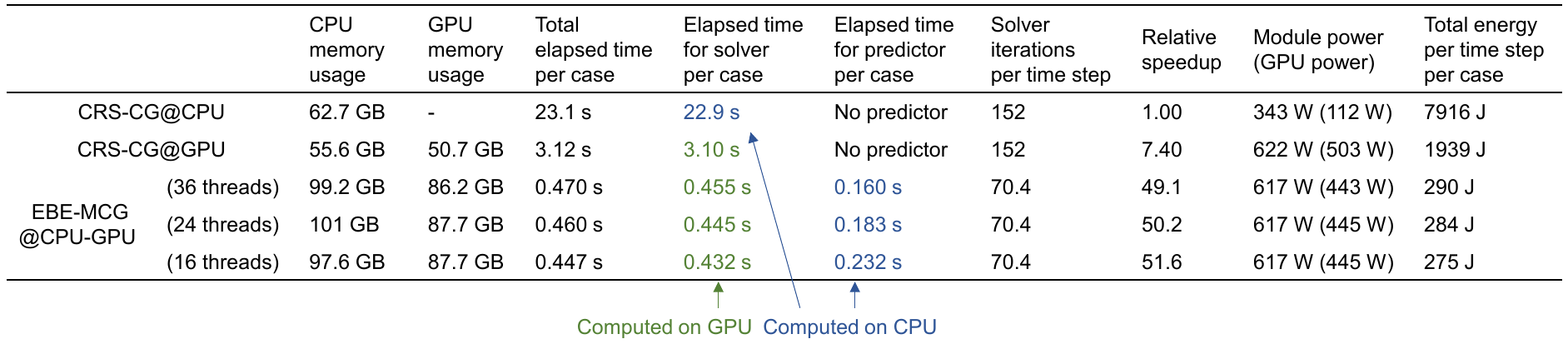}
\end{center}
\end{table}

First, we measure the performance on one Alps node. The target region described in Section~\ref{sct3a} is doubled in the $x, y$ directions, and its response under random impulse waves is computed using the 4 modules by MPI communication. First, we compare the performance of the conventional CRS-CG method (Table~\ref{performanceALPS1}). Compared to the single-GH200 node, CRS-CG@CPU is faster on Alps owing to the increased CPU memory bandwidth; additionally, the elapsed time of CRS-CG@GPU is slightly longer than on a single-GH200 node because of the power cap restricting GPU clock frequency.

We then measure the performance of the proposed method on a single Alps node (Table~\ref{performanceALPS1}). Whereas 32 time-steps could be stored for the predictor in the 480 GB CPU memory of the single-GH200 node, only 11 time-steps could be stored in the 128 GB CPU memory of Alps. On a single-GH200 node, the average number of iterations was 68.8, whereas on Alps, it was increased to 70.4 due to the reduction in predictor accuracy. Nevertheless, the number of iterations is reduced by a factor of 2.16 compared to the conventional methods using the Adams-Bashforth method; furthermore, with the performance improvement by the EBE-kernel with multiple right-hand sides results in a 49.1-fold speedup from CRS-CG@CPU and a 6.64-fold speedup from CRS-CG@GPU.

In systems where the CPU and GPU cannot run at high loads simultaneously owing to power constraints, the overall efficiency of computations can be improved by adjusting the ratio of power allocated to predictor@CPU and solver@GPU. In previous measurements, the predictor was computed using 36 CPU cores per process (corresponding to all available CPU cores). By reducing the number of threads from 36 to 24 or 16, we attempt to reduce the CPU power and increase the power allocated to the GPU. As the overall execution time is determined by the execution time of the solver part on the GPU side, the allocation of more power to the GPU is expected to reduce the overall execution time and improve energy efficiency. As shown in Table~\ref{performanceALPS1}, the time-averaged CPU and GPU power consumption do not significantly change regardless of the number of CPU threads; however, with the reduction in the number of CPU threads, the predictor time increases and power usage becomes more uniform. This led to a reduction in elapsed time, leading to a 51.6-fold speedup from CRS-CG@CPU and a 6.98-fold speedup from CRS-CG@GPU when using the optimal 16 OpenMP threads, and also shows a 28.8-fold and 7.04-fold improvement in terms of energy-to-solution. Thus, tailoring of the simulation parameters to the characteristics of the computer system allows efficient execution. Although the number of CPU cores used was adjusted in this case, there is potential for further improvement in execution time and energy performance when combined with other methods to constrain CPU power (e.g., frequency capping). 

\begin{figure}[tb]
\centering
\includegraphics[width=0.8\hsize]{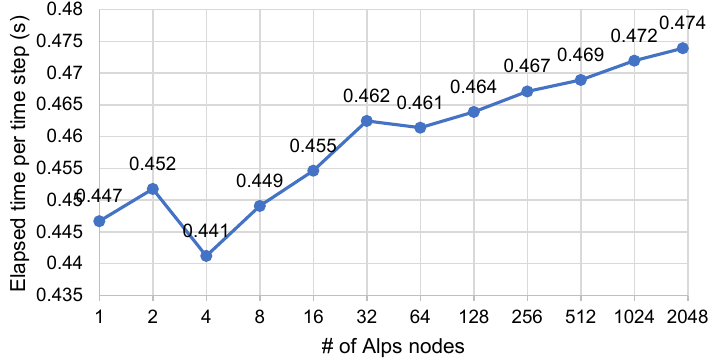}
\caption{Weak scaling of EBE-MCG@CPU-GPU on Alps. Time is shown for the average elapsed time per time step between 250--500th time-steps simulation per problem case. } 
\label{performanceALPS2}
\end{figure}

Finally, we measure the weak scaling performance of the proposed method on Alps. Here, the microtremors are calculated for a problem expanded in the $x,y$-directions from the problem run on one Alps node while ensuring the same problem size per compute node. Performance was measured from one node (4 GPUs) to 1,920 nodes (7,680 GPUs). The number of solver iterations per time step required to solve the problem is almost the same regardless of the problem size; thus, the computational load per compute node is also constant. As shown in Fig.~\ref{performanceALPS2}, the elapsed time is almost constant from 1 to 1,920 nodes (94.3\% weak scaling efficiency at 1,920 nodes). Thus, the proposed method, where data transfer is limited within the module of each compute node and scalable methods are used for both the predictor and solver, showed excellent scalability.

\section{Concluding Remarks}
\label{sct4}

To address the demand for the numerical solving of time-evolution PDE problems with guaranteed accuracy, we proposed a heterogeneous computing method in a general form. This method showed short time-to-solution and low energy-to-solution and exhibited the same accuracy as conventional equation-based modeling methods. While such analysis is usually computed using only the CPU or GPU, the proposed method combines a data-driven method, which takes advantage of the large CPU memory to improve energy-to-solution, with an equation-based modeling method, which takes advantage of the fast GPU computing, and is synchronized through a high-speed interconnect between the CPU and GPU. Although this method is somewhat complicated, we showed that it may be made portable using directive-based parallel programming models, enabling its application to the solution of actual problems while still yielding significantly better time-to-solution and energy-to-solution than the conventional method. Specifically, on a single-GH200 node, the time-to-solution was reduced by 86.4-fold compared to the baseline method run only on the CPU, and 8.67-fold compared to that run only on GPU. Furthermore, the energy-to-solution was reduced by 32.2 times (from 9944~J to 309~J) when compared to using only the CPU and reduced by 7.01 times (from 2163~J to 309~J) when compared to using only the GPU. On Alps, the proposed method showed 51.6-fold and 6.98-fold higher computation speed than the baseline method run only on CPU and only on GPU, respectively. Furthermore, a high weak scaling efficiency of 94.3\% up to 1,920 compute nodes was observed. Thus, a heterogeneous computing method, such as the proposed method, showing improved time-to-solution and energy-to-solution owing to the use of the heterogeneous environment can be realized by incorporating various methods. The method proposed herein is based on a general computation method to clearly show the performance improvement; however, it could be based on more sophisticated methods (e.g., solvers with improved convergence), which is expected to have even better performance. The time-to-solution and energy-to-solution improvements reported herein were realized through the use of directive-based parallel programming models, indicating that directives are effective in heterogeneous computing environments and are expected to be useful in future developments. Implementing this method on a different platform would demonstrate portability and provide opportunity to understand sensitivities to the relevant architectural features, e.g., CPU memory, CPU-GPU bandwidth, and GPU throughput. Such studies will be performed in future work. Furthermore, performance measurements using other programming models (e.g., OpenMP-based GPU offloading, Kokkos) is another future work when running on other computer architectures.

\section*{Acknowledgment}
This work was supported by a grant from the Swiss National Supercomputing Centre (CSCS) on Alps. We thank Yukihiko Hirano (NVIDIA) for coordination of the collaborative research project. This work was supported by JSPS KAKENHI Grant Numbers 23H00213, 22K18823. This work was supported by JST SPRING, Grant Number JPMJSP2108.


\begin{thebibliography}{100}

\bibitem{GH200}
NVIDIA GH200 Grace Hopper Superchip Architecture [Online]. 
\url{https://resources.nvidia.com/en-us-grace-cpu/nvidia-grace-hopper}
\url{https://docs.nvidia.com/gh200-superchip-benchmark-guide.pdf}

\bibitem{Gen5}
PCIe 5.0 specification [Online]. \url{https://pcisig.com/}

\bibitem{CRS-PCG-BJ}
Y. Saad, 2003, Iterative methods for sparse linear systems (2nd ed.), SIAM.

\bibitem{WACCPD2022}
R. Kusakabe, K. Fujita, T. Ichimura, M. Hori, and M. Lalith, GPU-Accelerated Sparse Matrix Vector Product based on Element-by-Element Method for Unstructured FEM using OpenACC, 2022 Workshop on Accelerator Programming Using Directives (WACCPD), Dallas, TX, USA, 2022, pp. 52--61.

\bibitem{EBE-FEM}
J. M. Winget and T. J. Hughes, Solution algorithms for nonlinear transient heat conduction analysis employing element-by-element iterative strategies, Computer Methods in Applied Mechanics and Engineering, 52, pp. 711-815, 1985.

\bibitem{HPCAsia2022}
T. Ichimura, K. Fujita, K. Koyama, R. Kusakabe, Y. Kikuchi, T. Hori, M. Hori, L. Maddegedara, N. Ohi, T. Nishiki, H. Inoue, K. Minami, S. Nishizawa, M. Tsuji, and N. Ueda, 152K-computer-node parallel scalable implicit solver for dynamic nonlinear earthquake simulation. In International Conference on High Performance Computing in Asia-Pacific Region (HPC Asia '22). Association for Computing Machinery, New York, NY, USA, 18--29, 2022

\bibitem{DMD}
P. J. Schmid, Dynamic mode decomposition of numerical and experimental data, Journal of Fluid Mechanics, vol. 656, pp. 5--28, 2010.

\bibitem{newmark}
N. M. Newmark, A method of computation for structural dynamics, Journal of the Engineering Mechanics Division, 85 (EM3), pp. 67-94, 1959.

\bibitem{FDD}
R. Brincker, L. Zhang, and P. Andersen, Modal identification of output-only systems using frequency domain decomposition. Smart Materials and Structures, 10(3), 441-445, 2001.

\bibitem{SCALA2022}
K. Fujita, S. Murakami, T. Ichimura, T. Hori, M. Hori, L. Maddegedara, and N. Ueda, Scalable Finite-Element Viscoelastic Crustal Deformation Analysis Accelerated with Data-Driven Method, 2022 IEEE/ACM Workshop on Latest Advances in Scalable Algorithms for Large-Scale Heterogeneous Systems (ScalAH), Dallas, TX, USA, 2022, pp. 18--25.

\bibitem{A64FX}
T. Yoshida, Fujitsu High Performance CPU for the Post-K Computer, IEEE Hot Chips: A Symposium on High Performance Chips, 2018.

\bibitem{Fugaku}
Supercomputer Fugaku, RIKEN Center for Computational Science [Online]. \url{https://www.r-ccs.riken.jp/en/fugaku/}

\bibitem{metis}
METIS - Serial Graph Partitioning and Fill-reducing Matrix Ordering [Online]. \url{https://github.com/KarypisLab/METIS}

\bibitem{openMP}
OpenMP [Online]. \url{https://www.openmp.org/}

\bibitem{openACC}
OpenACC  [Online]. \url{https://www.openacc.org/}

\bibitem{MPI}
MPI Forum [Online]. \url{https://www.mpi-forum.org/}

\bibitem{GPUDirect}
NVIDIA. (2023). GPUDirect Technology Overview. [Online]. \url{https://developer.nvidia.com/gpudirect}

\bibitem{ALPS}
CSCS Alps [Online]. \url{https://www.cscs.ch/computers/alps}

\bibitem{cusparse}
NVIDIA cusparse library [Online]. \url{https://docs.nvidia.com/cuda/cusparse/index.html}

\bibitem{nvidia-smi}
NVIDIA. (2023). nvidia-smi Documentation. [Online]. \url{https://developer.nvidia.com/system-management-interface}

\end{thebibliography}
\end{document}